\documentclass{aa}  

\usepackage{graphicx}
\usepackage{txfonts}
\usepackage{natbib} 
\bibpunct{(}{)}{;}{a}{}{,} 

\usepackage{hyperref}   
\hypersetup{colorlinks=true,linkcolor=blue,citecolor=blue,filecolor=blue,urlcolor=blue}

\begin{document}

   \title{Stellar model tests and age determination for RGB stars from the APO-K2 catalogue\footnote{Table 2 is only available in electronic form at the CDS via anonymous ftp to cdsarc.u-strasbg.fr (130.79.128.5) or via http://cdsweb.u-strasbg.fr/cgi-bin/qcat?J/A+A/}} 

   \subtitle{}
  \author{G. Valle \inst{1, 2}, M. Dell'Omodarme \inst{1}, P.G. Prada Moroni
        \inst{1,2}, S. Degl'Innocenti \inst{1,2} 
}
\titlerunning{Age determination for APO-K2 RGB stars}
\authorrunning{Valle, G. et al.}

\institute{
        Dipartimento di Fisica "Enrico Fermi'',
        Universit\`a di Pisa, Largo Pontecorvo 3, I-56127, Pisa, Italy
        \and
        INFN,
        Sezione di Pisa, Largo Pontecorvo 3, I-56127, Pisa, Italy
}


   \date{Received ; accepted }

  \abstract
{}
{By adopting the recently empirically derived dependence of $\alpha$-elements on $[\alpha/{\rm Fe}]$ instead of the conventionally applied uniform one, we tested the agreement between stellar model predictions and observations for red giant branch (RGB) stars in the APO-K2 catalogue.
We particularly focused on the biases in effective temperature scales and on the robustness of age estimations.}
{
We computed a grid of stellar models relying on the empirical scaling of $\alpha$-elements, investigating the offset in effective temperature  $\Delta T$ between these models and observations, using univariate analyses for both metallicity [Fe/H] and $[\alpha/{\rm Fe}]$. To account for potential confounding factors, we then employed a multivariate generalised additive model to study the dependence of $\Delta T$ on [Fe/H], $[\alpha/{\rm Fe}]$, $\log g$, and stellar mass.}
{
The initial analysis revealed a negligible trend of $\Delta T$ with [Fe/H], in contrast with previous works in the literature, which adopt a uniform relation between the various $\alpha$-elements and $[\alpha/{\rm Fe}]$. A slight $\Delta T$ difference of 25 K was detected between stars with high and low $\alpha$-enhancement. Our multivariate analysis reveals a dependence of $\Delta T$ on both [Fe/H] and $[\alpha/{\rm Fe}]$, and highlights a significant dependence on stellar mass. This suggests a discrepancy in how effective temperature scales with stellar mass in the models compared to observations.
Despite differences in assumed chemical composition, our analysis, through a fortunate cancellation effect, yields ages that are largely consistent with recent studies of the same sample. Notably, our analysis identifies a 6\% fraction of stars younger than 4 Ga within the high-$\alpha$ population. 
However, our analysis of the [C/N] ratio supports the possible origin of the these stars as a result of mergers or mass transfer events.
}
{} 

   \maketitle

\section{Introduction}\label{sec:intro}

The relevance of red giant branch (RGB) stars with precise observations  for Galactic archaeology investigations is widely recognised \citep[e.g.][]{Miglio2013b,   Stello2015, Miglio2017, Stasik2024}. In this respect, the availability of catalogues like APOGEE-Kepler  \citep{Pinsonneault2014} allows us  to obtain grid-based estimates of mass, radius, and age for thousands of stars in the RGB phase \citep[e.g.][]{Martig2015, Miglio2021, Warfield2021}.
However, unlike other stellar parameters, stellar ages cannot be measured directly and the only way to estimate them is by comparison to computed stellar models. 
It is therefore clear that these estimates are as reliable as the assumptions about the input physics and chemical abundances adopted in model computations. Moreover, age estimates are possibly affected by the presence of unknown biases, which may be different for different stellar evolutionary phases, observational uncertainties, and systematic errors \citep[e.g.][]{Gai2011, Basu2012, eta, Martig2015, smallsep}. For RGB stars, the effective temperature plays a crucial role in estimating stellar properties, such as age, using grid-based approaches \citep[see e.g.][]{Basu2010, Pinsonneault2018}. 

In this phase, theoretical effective temperature is intrinsically linked to the modelling of the superadiabatic convective temperature gradients in stellar evolution models. These models are almost universally based on
the simple, local formalism provided by the mixing length
theory \citep[MLT,][]{bohmvitense58}, which nevertheless has limitations in capturing the complex thermal structure of superadiabatic layers
\citep[see][for a review]{Joyce2023}. 
Choosing the right value for the free parameter of the method, $\alpha_{\rm MLT}$, plays a crucial role. 
Traditionally, $\alpha_{\rm MLT}$ is  calibrated by matching the properties of the Sun at the solar age. However, assuming a universal $\alpha_{\rm MLT}$ for all stars regardless of mass, evolutionary stage, and chemical composition is possibly unrealistic.
This uncertainty affects determinations of the effective temperature of  RGB stars  in a relevant way, because these calculations are very sensitive to the treatment
of the superadiabatic layers \citep[e.g.][and references therein]{Straniero1991, Salaris1996, Vandenberg1996}. 
On the other hand, the observational determination of the effective temperature of RGB stars is also problematic, and  systematic biases of 50-70 K exist between different surveys \citep[e.g.][]{Hegedus2023, Yu2023}.

The existence of discrepancies between observed and predicted values led to the hypothesis that there may be a dependence of $\alpha_{\rm MLT}$ on metallicity \citep{Tayar2017, Salaris2018}, casting doubts on the accuracy of  stellar
models for objects
at high $[\alpha/{\rm Fe}]$. However, as discussed by \citet{ML}, other uncontrolled factors, such as a mismatch in the solar reference mixture or in the initial helium abundance, may act as confounding variables in the analysis. 
An opportunity to gain further insight into this topic was offered by the recent release of the APO-K2 catalogue \citep{Stasik2024}, which contains  high-precision data for 7,673 RGB and red clump stars, combining  spectroscopic \citep[APOGEE DR17,][]{Abdurrouf2022}, asteroseismic \citep[K2-GAP,][]{Stello2015}, and astrometric \citep[Gaia EDR3,][]{Gaia2021} data. This catalogue significantly improved the sampling of the region at high $\alpha$ enhancement. 
Leveraging this large data set, \citet{Warfield2024} presented age estimates for the RGB stars in the catalogue, but many aspects remain unexplored.  
Therefore,  here we revisit the determination of the stellar ages and explore the strength of the agreement between predicted and observed effective temperatures. Our aims are twofold. First, we aim to check the reliability of the assumptions in the computations of 
stellar models and to verify the presence of systematic trends. Second, we aim to obtain robust estimates of the catalogue stellar ages and to compare them with existing literature. 

A key difference between this and previous studies is that the stellar models adopted in the present analysis account for the individual observed scaling of chemical elements with $[\alpha/{\rm Fe}]$ obtained from the analysis by \citet{Valle2024cno}\defcitealias{Valle2024cno}{V24} (hereafter \citetalias{Valle2024cno}). 
A noticeable disagreement was reported in \citetalias{Valle2024cno} in the [(C+N+O)/Fe] -- $[\alpha/{\rm Fe}]$ relationship with respect to the simple $\alpha$-enhancement scenario. In particular, the evolution of oxygen with $[\alpha/{\rm Fe}]$ was found to be significantly steeper than those of other elements, as already reported in the literature \citep[e.g.][]{Bensby2005, Nissen2014, deLis2015, Amarsi2019}.  This discrepancy could potentially affect the agreement between the model effective temperatures and the observed values, with a consequent bias in the estimated stellar ages.
While \citetalias{Valle2024cno} offered some reassurance by testing a limited scenario with only oxygen enrichment, we present a more comprehensive test in this paper. This allows us to assess the sensitivity of the results to this refined chemical treatment.

\section{Methods}\label{sec:method}
\subsection{Stellar models grid}

The grid of stellar evolutionary models was calculated for the mass range 0.8 to 1.9 $M_{\sun}$, spanning the evolutionary stages from the pre-main sequence to the tip of the RGB.
The initial metallicity [Fe/H] was varied from $-1.5$ dex to 0.4 dex with
a step of 0.025 dex. 
We adopted the solar heavy-element mixture by \citet{AGSS09}. 
The initial helium abundance was set from the commonly used linear relation $Y = Y_p+\frac{\Delta Y}{\Delta Z} Z$
with the primordial helium abundance  $Y_p = 0.2471$ from \citet{Planck2020}.
The helium-to-metal enrichment ratio $\Delta Y/\Delta Z$ was set
to 2.0. $\alpha$ enhancement was considered for $\alpha$ elements and Al. For every element X, we adapted a linear model to the evolution [X/Fe] with $[\alpha/{\rm Fe}]$ as reported in \citetalias{Valle2024cno}. The values $c_X$  of these slopes (Table~\ref{tab:ele-a}) were used to describe the variation of individual elements with $[\alpha/{\rm Fe}]$: [X/Fe] = $c_X \times [\alpha/{\rm Fe}]$.  Stellar models were computed for $[\alpha/{\rm Fe}]$ from $-0.1$ to 0.4 with a step of 0.1 dex. 

\begin{table}
        \caption{Dependence of the different element  abundances [X/Fe] on $[\alpha/{\rm Fe}]$. }\label{tab:ele-a}
        \centering
        \begin{tabular}{lc}
                \hline\hline
Element & Coefficient\\
\hline
O  &  1.31 \\
Ne & 1.00 \\
Mg & 1.14 \\
Al & 0.93 \\
Si & 0.88 \\
S  & 1.00\\
Ar & 1.00\\
Ca & 0.78 \\
Ti & 1.00\\
\hline
        \end{tabular}
\end{table}

Models were computed with the FRANEC code. The input physics was the same
 as that adopted to compute the Pisa Stellar
Evolution Data Base\footnote{\url{http://astro.df.unipi.it/stellar-models/}} 
for low-mass stars \citep{database2012}, apart from the following differences. The outer boundary conditions  were set by the \citet{Vernazza1981} solar semi-empirical $T(\tau)$, which well approximate the results obtained using the hydro-calibrated $T(\tau)$ \citep{Salaris2015, Salaris2018}.
The models were computed
assuming the solar-scaled mixing-length parameter $\alpha_{\rm
        ml} = 1.91$.
A moderate mass-loss was assumed according to the parametrisation by \citet{Reimers1975}, with an efficiency of $\eta = 0.2$.
High-temperature ($T > 10,000$ K) radiative opacities were taken from the OPAL group
\citep{rogers1996}\footnote{\url{http://opalopacity.llnl.gov/}},
whereas for lower temperatures the code adopts molecular opacities by
\citet{Marigo2022}\footnote{\url{http://stev.oapd.inaf.it/cgi-bin/aesopus}}. Both high- and low-temperature opacity tables account for the metal distributions adopted in the computations.

Raw stellar evolutionary tracks were reduced to a set of tracks with the same number of homologous points according to the evolutionary phase.
Details about the reduction procedure are reported in the Appendix of \citet{incertezze1}.  Only models in RGB with $1.4 \leq \log g \leq 3.3$ and an age of below 14 Ga were retained in the final grid.

\subsection{Effective temperature discrepancies}\label{sec:fit-teff}

For each star in the sample, we compared the observed effective temperature $T_{\rm eff, obs}$ with a theoretical value $T_{\rm eff, mod}$ estimated by linearly interpolating over the grid of stellar models. The interpolation was carried out in mass, $\log g$, $[\alpha/{\rm Fe}]$, and [Fe/H]. 
The difference between the observed and estimated temperatures defines the effective temperature offset, $\Delta T = T_{\rm eff, obs} - T_{\rm eff, mod}$.
Stellar masses used in the process were derived from asteroseismic scaling relations and sourced from the APO-K2 catalogue. These mass estimates apply a correction to the face $\Delta \nu$ values used in the scaling relations, with the coefficient $f_{\Delta \nu}$ derived from \citet{Sharma2016}. It is therefore important to acknowledge that a possible uncontrolled systematic bias present in these estimates may affect the obtained results.
Correction factor variations of up to 2\% are reported in the literature along with possible systematic trends related to stellar mass, metallicity, and $\log g$ 
\citep{Li2022, Li2023}. 
While a thorough investigation of these effects across different correction methods would be desirable, it falls outside the scope of this paper.

Our approach is the same as that adopted by  \citet{Tayar2017} and \citet{Salaris2018}, who performed the analysis using different stellar models over a  smaller sample of RGB stars. 
Therefore, a direct comparison of the results could deepen our understanding of the discrepancies between stellar models and observations, and particularly with regard to how these discrepancies vary with different parameters such as [Fe/H] or $[\alpha/{\rm Fe}]$.

\subsection{Statistical models}\label{sec:fit-stat}

Previous studies \citep[e.g.][]{Tayar2017, Salaris2018, ML} have explored the relationship between effective temperature offset $\Delta T$ and metallicity [Fe/H] or $\alpha$ enhancement. However, they might not fully account for biases arising from differences in stellar populations across the [Fe/H]--$[\alpha/{\rm Fe}]$ plane. For instance, variations in stellar mass distributions at different $[\alpha/{\rm Fe}]$ values could distort the results.
To address this, a multivariate approach is needed that considers the interplay between multiple variables simultaneously. 
This is particularly important because investigations into marginalised trends of $\Delta T$ can lead to misleading conclusions given the hidden influences from other neglected factors. In our analysis, we use a generalised additive model \citep[GAM,][]{Hastie1986, Wood2017} to capture potential non-linear relationships between $\Delta T$ and various observational parameters.

GAMs extend generalised linear models by allowing the response variable $Y$ to depend on smooth, unknown functions of the predictor variables $x$. These functions do not need a predefined form, and the optimal level of smoothness for each is automatically determined by the model.
The general GAM  form is: 
\begin{equation}
        g(\mu_i) = A_i \theta + f_1(S_1(x)) + f_2(S_2(x)) + f_3(S_3(x) ) + \ldots  
\end{equation}
where $g(.)$ is a link function, $\mu_i  = \mathbb{E}(Y_i)$ are the expected values of $Y_i$, and $Y_i \sim EF(\mu_i , \phi)$, where $EF(\mu_i , \phi)$ is an exponential family distribution with mean $\mu_i$ and scale parameter $\phi$. Also, $A_i$ is a row
of the model matrix for any parametric model components, $\theta$ is the corresponding parameter vector, and $f_j(.)$ are smooth functions of the different subsets $S_j$ of the covariates $x$. 
The smoothing functions $f_j(.)$ are then expanded over a basis of functions, which are completely known. If $b_j(x)$ is the $j$-th such basis function, then $f(.)$ are
assumed to have a representation
\begin{equation}
f(x) = \sum_{j=1}^k b_j(x) \; \beta_j
,\end{equation}
where  $\beta_j$ are unknown parameters and $k$ is the basis space dimension.
We estimated the smooth functions in the model using penalised regression methods. The degree of smoothness was automatically determined from the data using the generalised cross-validation criterion as described in \citet{Wood2017}. We used the {\it gam} function from the {\it mgcv} package in R 4.3.2 \citep{R} to fit the model. This function employs penalised regression splines with basis functions optimised for the chosen number of basis functions \citep{Wood2017}.

The fitted model was
\begin{equation}
        \Delta T =  f_1({\rm[Fe/H]} , [\alpha/{\rm Fe}] , \log g) + f_2(M) + [\alpha/{\rm Fe}]*{\rm[Fe/H]}*M \label{eq:gam}
,\end{equation}
where the asterisk in the model specification indicates that we fit not only the individual terms
but also their interactions, capturing potential combined effects. This results in a model with three components: a parametric component for the interaction term 
($[\alpha/{\rm Fe}]*{\rm[Fe/H]}*M$), a smoothing function of the mass $M$, and a multivariate smoothing function of [Fe/H], $[\alpha/{\rm Fe}],$ and $\log g$.
The adequacy of the chosen base dimensions for the smoothing functions was verified using the $k'$ test \citep{Wood2017}. 
 
To ensure the robustness of our conclusions, we additionally fit an alternative model using projection pursuit regression \citep{venables2002modern, simar}. However, both models yielded similar predictions. Therefore, we only present the results based on the GAM model.

\subsection{Age-fitting technique}\label{sec:fit-method}

 The analysis was conducted adopting the SCEPtER pipeline\footnote{Publicly available on CRAN: \url{http://CRAN.R-project.org/package=SCEPtER}}, a well-tested technique for fitting single and binary systems
\citep[e.g.][]{scepter1,eta, Valle2023a}. 
The pipeline estimates the stellar age adopting a grid maximum likelihood  approach, relying on the observed quantities $o \equiv \{T_{\rm eff}, {\rm [Fe/H]}, \Delta \nu, \nu_{\rm max}\}$. 
For every star, a fitting grid with $[\alpha/{\rm Fe}]$ equal to the observed values is obtained by means of linear interpolation. Then, for every $j$-th point in the fitting grid, a likelihood estimate is obtained
\begin{equation}
        {\cal L}_j = \left( \prod_{i=1}^n \frac{1}{\sqrt{2 \pi}
                \sigma_i} \right) 
        \times \exp \left( -\frac{\chi^2}{2} \right)
        \label{eq:lik}
        ,\end{equation}
\begin{equation}
        \chi^2 = \sum_{i=1}^n \left( \frac{o_i -
                g_i^j}{\sigma_i} \right)^2
        \label{eq:chi2},
\end{equation}
where $o_i$ are the $n$ observational constraints, $g_i^j$ are the $j$-th grid point corresponding values, and $\sigma_i$ are the observational uncertainties. The estimated stellar ages were obtained
by averaging the  age of all the models with a likelihood
of greater than $0.95 \times L_{\rm max}$\footnote{
The likelihood threshold is set for computational efficiency. Negligible differences arise using a weighted mean where each model's weight reflects its likelihood.}.
The confidence interval for these estimates was obtained by means of a Monte Carlo simulation. We repeated the estimation over a generated sample of size $n = 1,000$ for every star. These synthetic stars were obtained  by means of Gaussian perturbations around their observed values, adopting a diagonal covariance matrix  $\Sigma = {\rm diag}(\sigma^2)$. 
The median of the $n$ ages was taken as the best estimate of the true values;
the 16th and 84th percentiles of the $n$ values were adopted as a $1 \sigma$
confidence interval. 

Nevertheless, concerns have been raised regarding the use of effective temperature as a constraint in fitting procedures due to both observational and theoretical uncertainties \citep[see e.g.][]{Martig2015, Warfield2021, Vincenzo2021}.
Biases in effective temperature values relative to the fitting grid can significantly impact estimated ages through their influence on mass, which directly affects age determination.
To deal with this issue, we adopted $T_{\rm eff, mod}$, the theoretical interpolated effective temperatures, as observational constraints in the fit \citep[][]{Martig2015, Warfield2021, Warfield2024}.

\section{Test of stellar models}\label{sec:results}

\subsection{Marginalised univariate investigations}

\begin{figure*}
        \centering
        \includegraphics[height=8.4cm,angle=0]{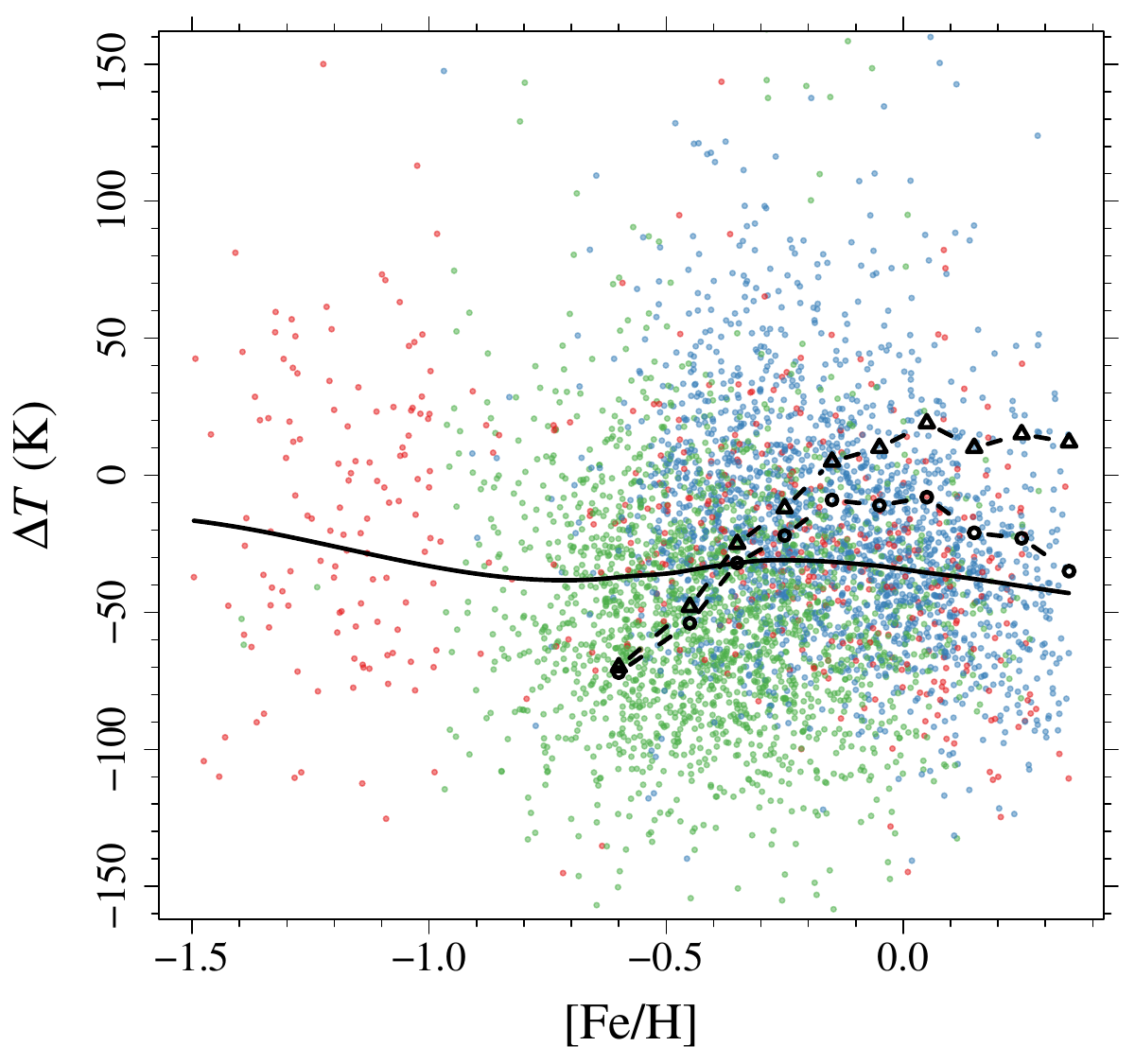}
        \includegraphics[height=8.4cm,angle=0]{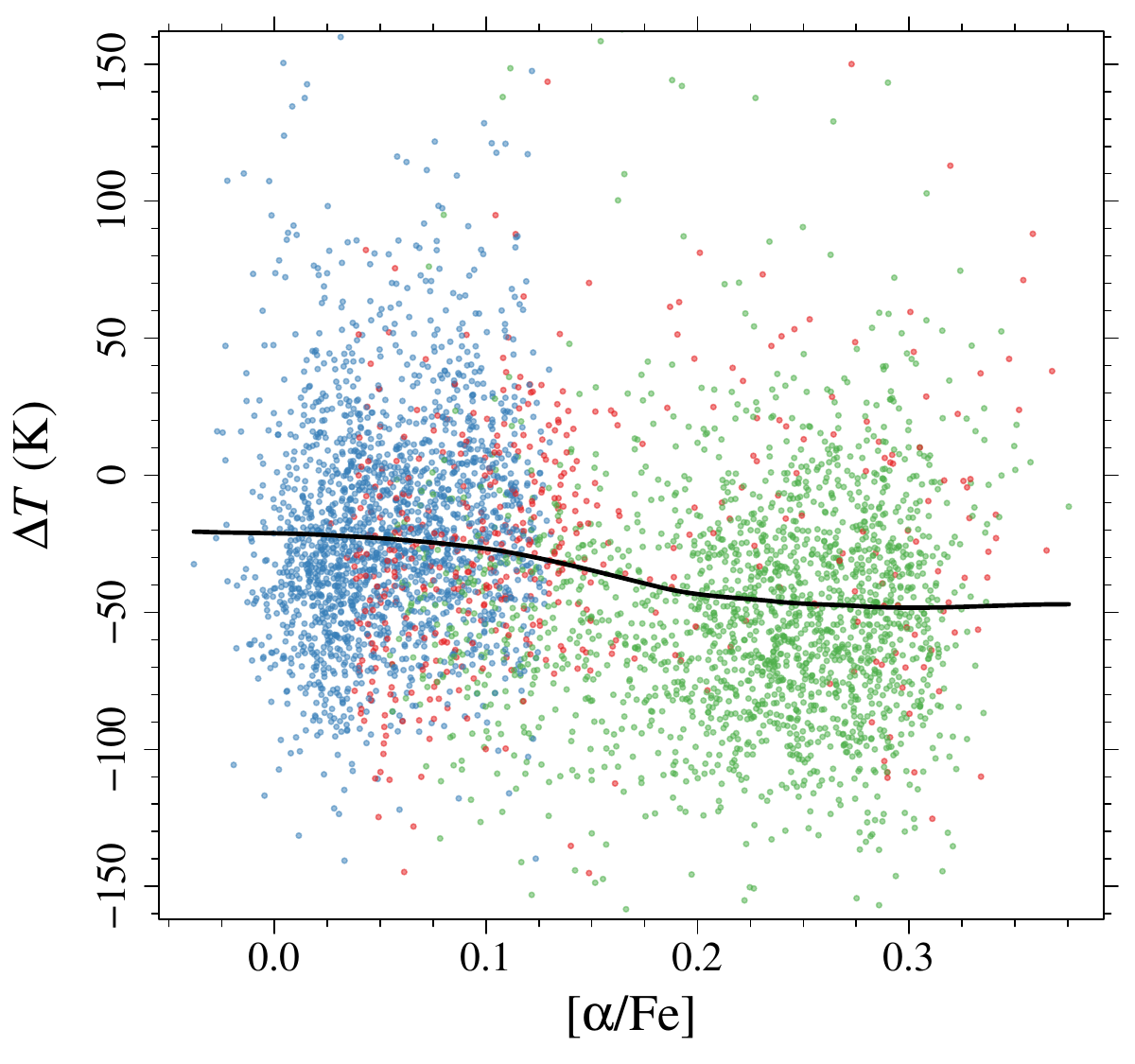}
        \caption{Effective temperature difference between observations and models as a function of different observational parameters. ({\it Left}): 
                $\Delta T$ vs [Fe/H]. Colours identify stars as in the APO-K2 catalogue (green = high $\alpha$, blue = low $\alpha$, red = transition). The black solid line represent a loess smoothing 
                of the data. Open circles show the trend reported in \citet{Salaris2018}, while open triangles show the trend by \citet{Tayar2017}, shifted downward by 100 K.  ({\it Right}): Same as in the left panel but as a function of $[\alpha/{\rm Fe}]$.    }
        \label{fig:dt-feh+alpha}
\end{figure*}

Figure~\ref{fig:dt-feh+alpha} presents the difference between observed and predicted effective temperatures $\Delta T$ as a function of metallicity [Fe/H] and  $[\alpha/{\rm Fe}]$.
Overall, the trend with [Fe/H] suggests a slight systematic bias in the models, with a median overprediction of temperature by about 35 K. This trend remains relatively flat across the metallicity range.
The left panel also shows the trends reported by \citet{Tayar2017} and \citet{Salaris2018} for comparison. The values from \citet{Tayar2017} were rebinned as in \citet{Salaris2018} and shifted downward by 100 K for an easy visual comparison.
Our results differ significantly from those of the previous studies. The stellar evolution code adopted here is very similar to that by \citet{Salaris2018}, the main difference being the correction of the element trend with $[\alpha/{\rm Fe}]$. However, definitively attributing the differences in the $\Delta T$ agreement solely to this specific modification would be premature. 
As the following discussion shows, other factors like the underlying distributions of $[\alpha/{\rm Fe}]$, mass, and $\log g$ in stars across different [Fe/H] intervals could also influence this trend. Moreover, it is relevant to note that the analyses of \citet{Tayar2017}
 and \citet{Salaris2018}  were based upon APOGEE DR13 data, and several differences have arisen in the APOGEE fitting and calibration pipeline  since those studies \citep{Holtzman2018, Jonsson2020}. 
 The effective temperatures provided in the APOGEE DR17 data set used in this paper were calibrated based on the work by \citet{Gonzales2009}. This calibration successfully removed a significant dependence of the spectroscopic $T_{\rm eff}$ on the metallicity [Fe/H]. A comparison of these calibrated effective temperatures with those from other catalogues \citep{Hegedus2023} did not reveal any significant residual trends or biases, further supporting the robustness of the analysis presented here.

The right panel of Fig.~\ref{fig:dt-feh+alpha} presents $\Delta T$ values against $[\alpha/{\rm Fe}]$, highlighting the model's differing accuracy at low and high $\alpha$-values. Stars with high $\alpha$ enhancement exhibit a roughly 50 K disagreement with the model, while the discrepancy is only about 25 K for stars with low $\alpha$ enhancement. The difference in accuracy as a function of $[\alpha/{\rm Fe}]$ aligns with the finding by \citet{Salaris2018}, who reported improved agreement in $\Delta T$ when restricting to $[\alpha/{\rm Fe}] < 0.07$ stars. 
However, the same caveat as that mentioned for [Fe/H] trend regarding premature conclusions applies here as well. 
Interestingly, repeating the analysis with the uncalibrated APOGEE temperature \texttt{TEFF\_SPEC} revealed a significantly larger discrepancy of 56 K between the two groups. This discrepancy also showed a steep dependence of $\Delta T$ on $[\alpha/{\rm Fe}]$, suggesting that the residual difference between the low- and high-$\alpha$ populations might, at least partially, originate from an inhomogeneous calibration across the entire $[\alpha/{\rm Fe}]$ range.
 
\begin{figure}
        \centering
        \includegraphics[width=8.6cm,angle=0]{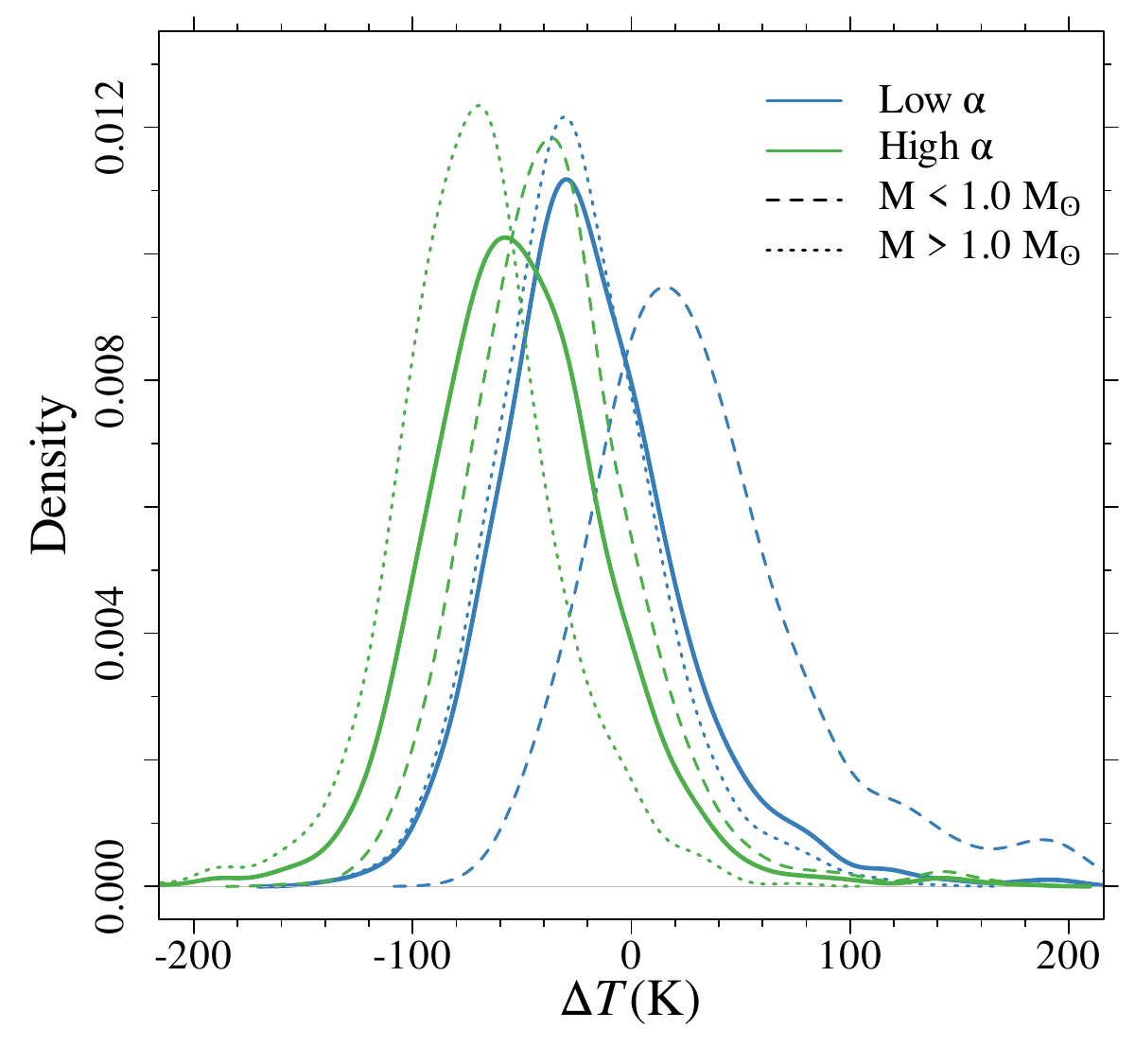}
        \caption{Kernel density estimator of the $\Delta T$ distribution in low- and high-$\alpha$ populations (solid lines). Dashed lines show the distributions restricted to $M < 1.0$ $M_{\sun}$, while dotted lines correspond to   $M > 1.0$ $M_{\sun}$ populations. }
        \label{fig:dt-dens}
\end{figure}

Figure~\ref{fig:dt-dens} displays the kernel density estimators of the differences in predicted effective temperature accuracy for high- and low-$\alpha$ populations. The figure also shows the kernel density estimator in the subpopulations obtained by separating stars according to their mass, adopting  1.0 $M_{\sun}$ for the cut.
For both populations, $\Delta T$ values are consistently higher by about 25 K in lower mass bins. This trend contradicts the findings of \citet{Tayar2017}, who reported a positive linear correlation (Pearson coefficient $r=0.12$) between mass and $\Delta T$. Our analysis yields a negative correlation ($r=-0.23$), but the strong nonlinearity in the data suggests this value may not be meaningful.
Furthermore, the analysed sample lacks homogeneity within the observational hyperspace, introducing potential biases due to uncontrolled confounding variables in marginalised analyses.
For instance, the observed mass dependence of $\Delta T$ is particularly relevant considering the significant mass differences between high- and low-$\alpha$ populations. The low-$\alpha$ group has a median mass of 1.18 $M_{\sun}$ (interquartile range [1.08, 1.30] $M_{\sun}$), while the high-$\alpha$ stars have a median of 0.98 $M_{\sun}$ (interquartile range [0.92, 1.06] $M_{\sun}$). This disparity could significantly bias the results of marginalised analyses presented earlier.

\subsection{Non-linear multivariate approach}

\begin{figure*}
        \centering
        \includegraphics[width=8.6cm,angle=0]{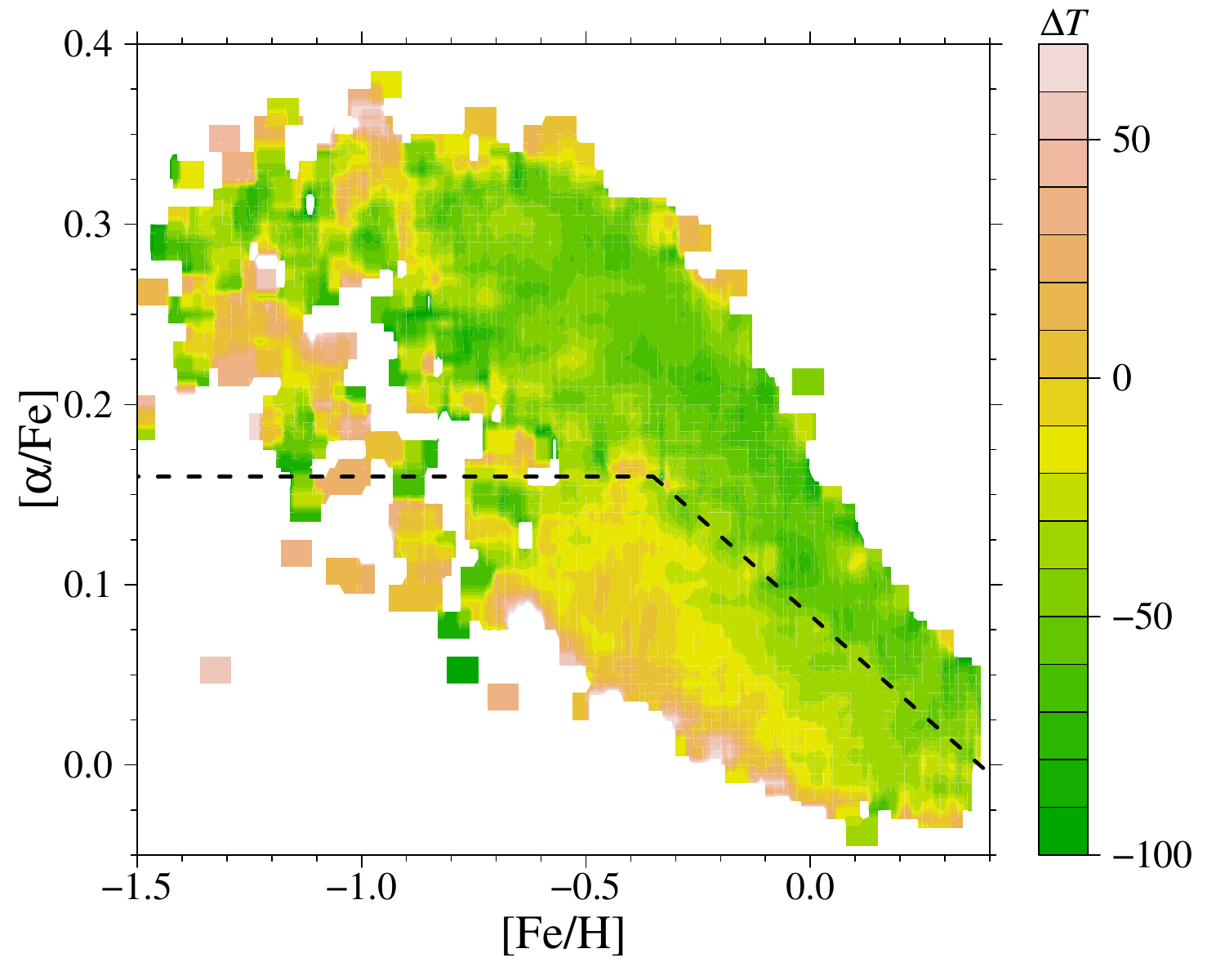}
        \includegraphics[width=8.6cm,angle=0]{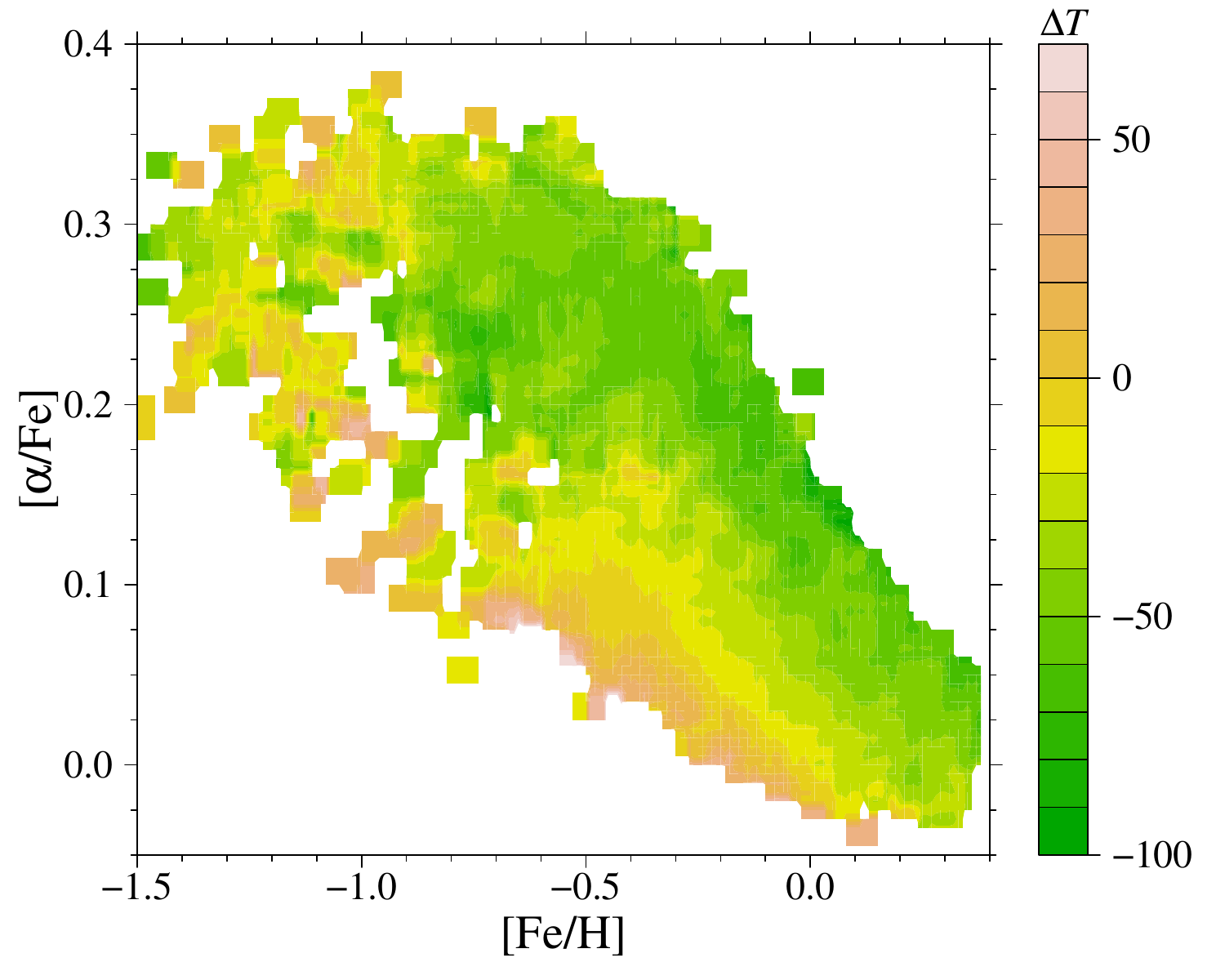}
        \caption{Two-dimensional plot of $\Delta T$ as a function of [Fe/H] and $[\alpha/{\rm Fe}]$. The black dashed line marks the separation adopted in the APO-K2 catalogue to define low- and high-$\alpha$ populations. ({\it Left}): 
                Map of the observed differences. ({\it Right}): Map of the differences as predicted by GAM model.   }
        \label{fig:dt-feh-alpha}
\end{figure*}

To deal with this fundamental question and disentangle the contributions from the different stellar observables to the $\Delta T$ value, a robust and multivariate approach is mandatory. We adopt a GAM for this analysis, as detailed in Section~\ref{sec:fit-stat}. This model class offers superior flexibility compared to linear models, enabling both accurate predictions and the avoidance of overfitting
\citep{Wood2017}. The adequacy of the fitted GAM, specified in Eq.~(\ref{eq:gam}), is evaluated in Appendix~\ref{app:gam}.

The first test of the model is its capability to reproduce the $\Delta T$ values across the $[\alpha/{\rm Fe}]$ versus [Fe/H] plane. In the left panel of Fig.~\ref{fig:dt-feh-alpha}, a filled contour plot of the $\Delta T$ in this plane is shown (a positive $\Delta T$ value means that the observed temperature is greater than the theoretical one). The plot was produced by computing the mean value of the neighbourhood stars $\Delta T$  over a regular grid.  A bidimensional rectangular kernel with bandwidths from  \citet{Sheathe1991} is adopted for the mean computation. It is apparent that models become hotter moving from the low- to the high-$\alpha$ groups, with a reduced  trend  within the two groups. The plot shows that the   $\Delta T$ values evolve mostly parallel to the separation line  adopted in the APO-K2 catalogue to identify low- and high-$\alpha$ populations. The
same plot is shown in the right panel of Fig.~\ref{fig:dt-feh-alpha}  but this time using $\Delta T$ values as predicted by GAM. While adjusting a few outliers in the real data, the model demonstrably aligns with observed trends, supporting the credibility of its predictions.

\begin{figure*}
        \centering
        \includegraphics[width=8.6cm,angle=0]{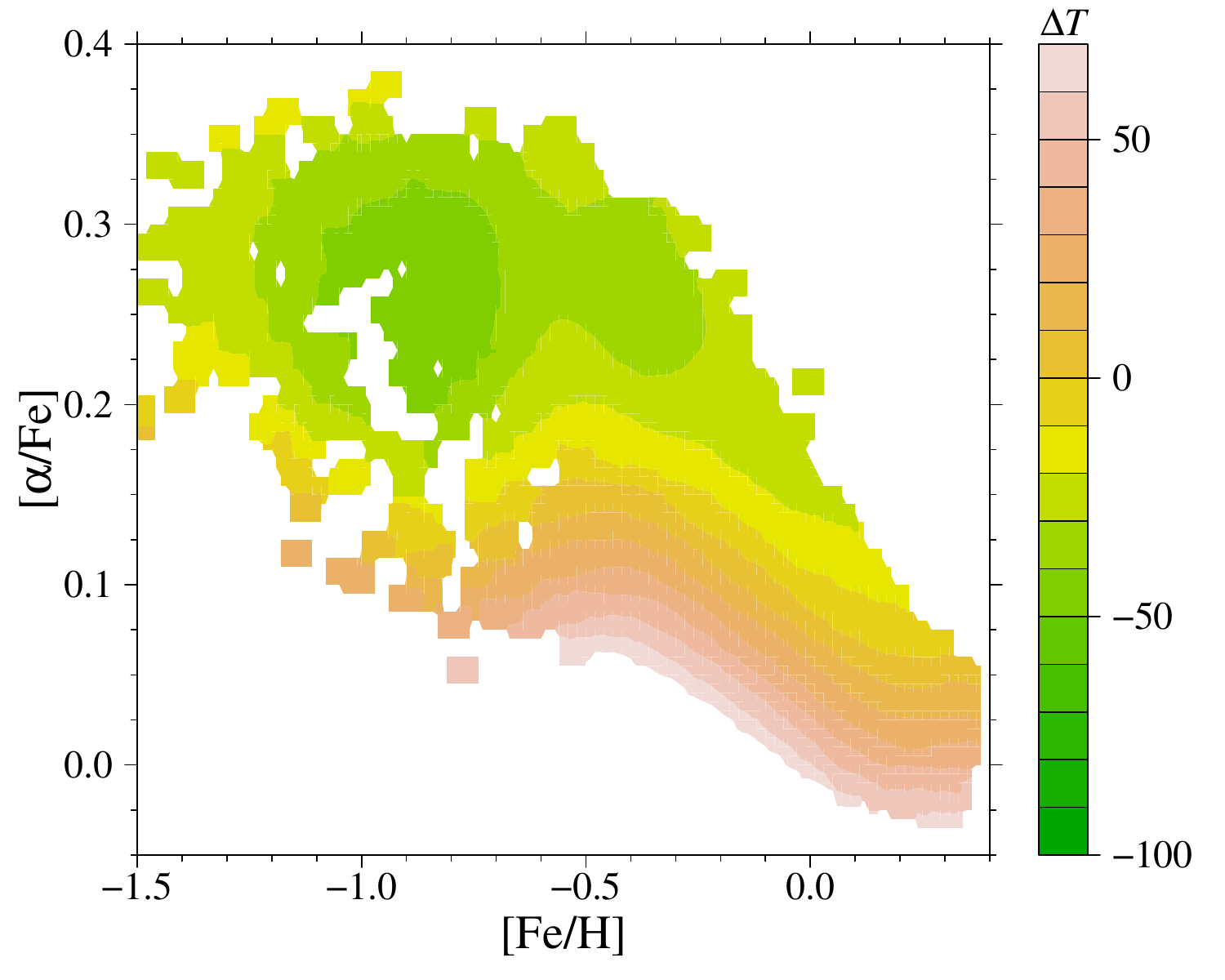}
        \includegraphics[width=8.6cm,angle=0]{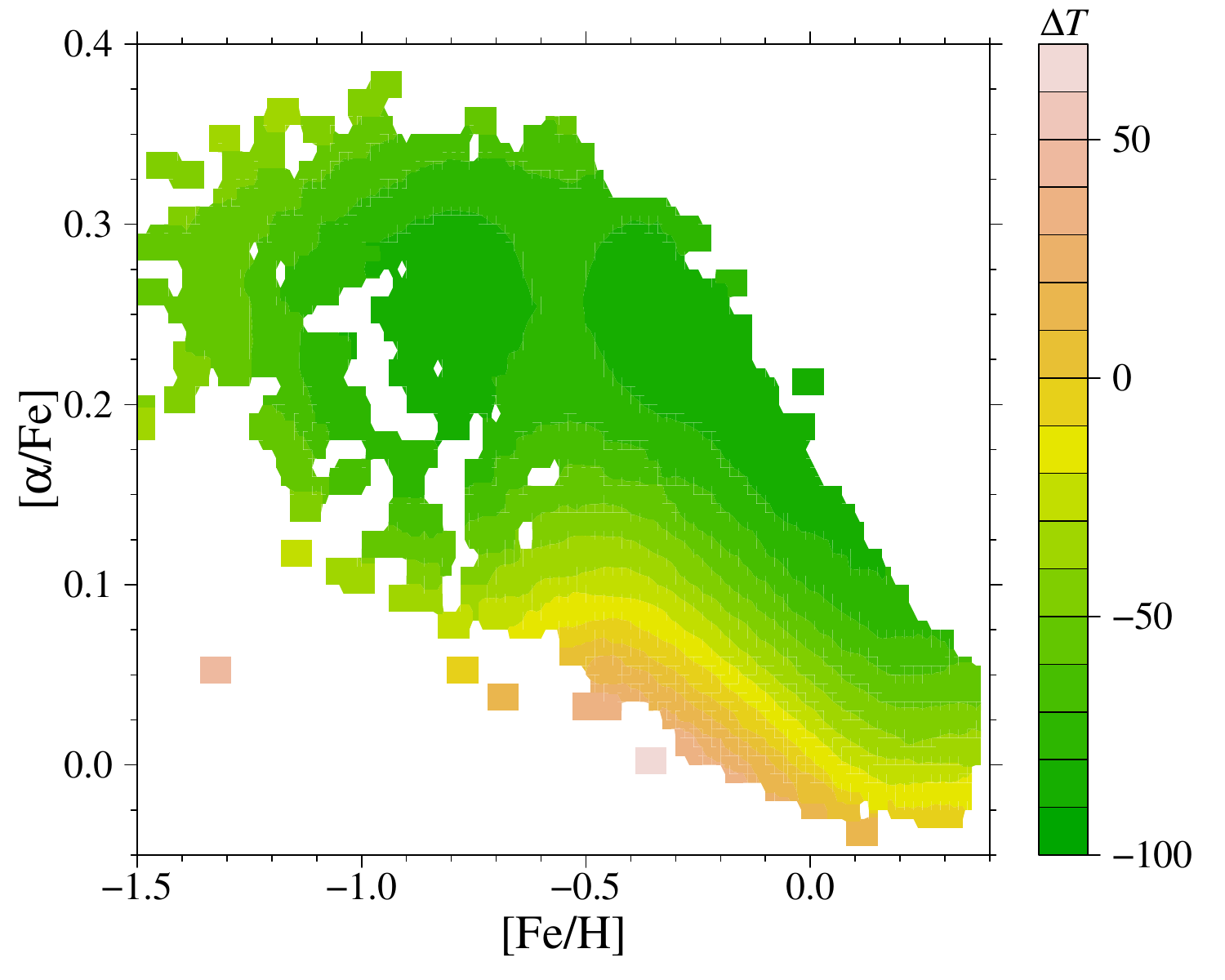}
        \caption{Two-dimensional plot of $\Delta T$ from GAM predictions as a function of [Fe/H] and $[\alpha/{\rm Fe}]$ when fixing the masses to different values. {\it Left}: Stellar masses set to 0.9 $M_{\sun}$. 
        {\it Right}:  Stellar masses set to 1.2 $M_{\sun}$. }
        \label{fig:dt-models}
\end{figure*}

The availability of a reliable statistical model allows  us to explore the individual contributions of the different stellar parameters to $\Delta T$.  The interpretation of the trend seen in Fig.~\ref{fig:dt-feh-alpha} is indeed complex because the mass distribution significantly differs as a function of [Fe/H] and $[\alpha/{\rm Fe}]$. A much clearer view would result from filtering out this confounding factor. Using the GAM to predict $\Delta T$ while 
assigning the same mass and $\log g$ value  to all the stars  
would reveal the actual trend of $\Delta T$ with [Fe/H] and $[\alpha/{\rm Fe}]$. Moreover, comparing the results produced by fixing the stellar mass to different values would
help us to clarify the trends in mass given [Fe/H] and $[\alpha/{\rm Fe}]$.
To this aim, Fig.~\ref{fig:dt-models} shows the maps obtained by fixing stellar masses to 0.9 and  1.2 $M_{\sun}$ and $\log g = 2.7$. 

The analysis of this figure reveals three key findings regarding the discrepancies between models and observations.
\begin{itemize}
\item 
The two maps exhibit notable variations in $\Delta T$ as a function of stellar mass for the same zone in the $[\alpha/{\rm Fe}]$ -- [Fe/H] plane. This suggests that the model--observation agreement is mass-dependent, implying that changes in effective temperature between models at different masses do not correctly reflect observed changes.
\item The $\Delta T$ offset evolves strongly when moving along the $[\alpha/{\rm Fe}]$ axis. At [Fe/H] = $-0.5,$ there is a difference of about 100 K in the effective temperature offset between the extreme $[\alpha/{\rm Fe}]$ zones. 
The $\Delta T$ dependence on  [Fe/H]  is weaker, with a typical maximum variation  of about one-half of that due to $[\alpha/{\rm Fe}]$. 
\item The $\Delta T$ 
offset evolves  nearly parallel to the line adopted in the separation of high- and low-$\alpha$ populations, in agreement with the evidence discussed above for the global map.
\end{itemize}

\begin{figure}
        \centering
        \includegraphics[width=8.6cm,angle=0]{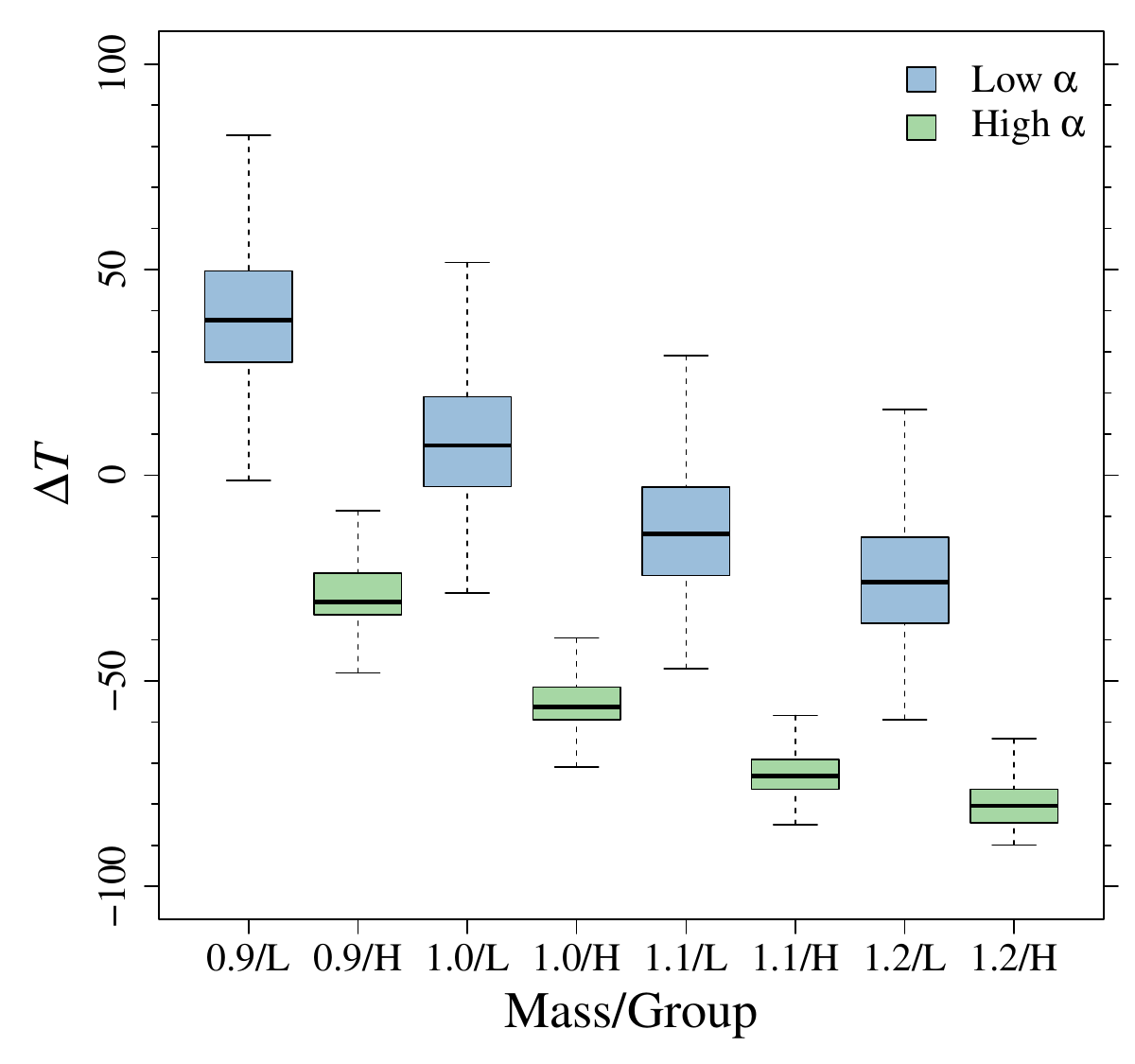}
        \caption{Boxplot of  $\Delta T$  as a function of the stellar mass and $\alpha$ enhancement group (L = low-$\alpha$, H = high-$\alpha$).}
        \label{fig:dT-boxplot}
\end{figure}

Figure~\ref{fig:dT-boxplot} further explores the dependence of $\Delta T$ on stellar mass.
The boxplots show the distribution of effective temperature errors predicted by the GAM model when all data set masses are fixed. Stellar masses in the range from 0.9 $M_{\sun}$ to 1.2 $M_{\sun}$ are considered. This reveals a clear mass dependence in model accuracy, with the median $\Delta T$ varying by approximately 60 K between the maximum and minimum considered masses.
The difference in effective temperature between low- and high-$\alpha$ populations slightly decreases with increasing mass from about 70 K to about 55 K. These values are significantly higher than the 25 K inferred from the analysis of the right panel in Fig.~\ref{fig:dt-feh+alpha}, highlighting the significance of mass as a confounding factor in univariate analyses.

\begin{figure}
        \centering
        \includegraphics[width=8.6cm,angle=0]{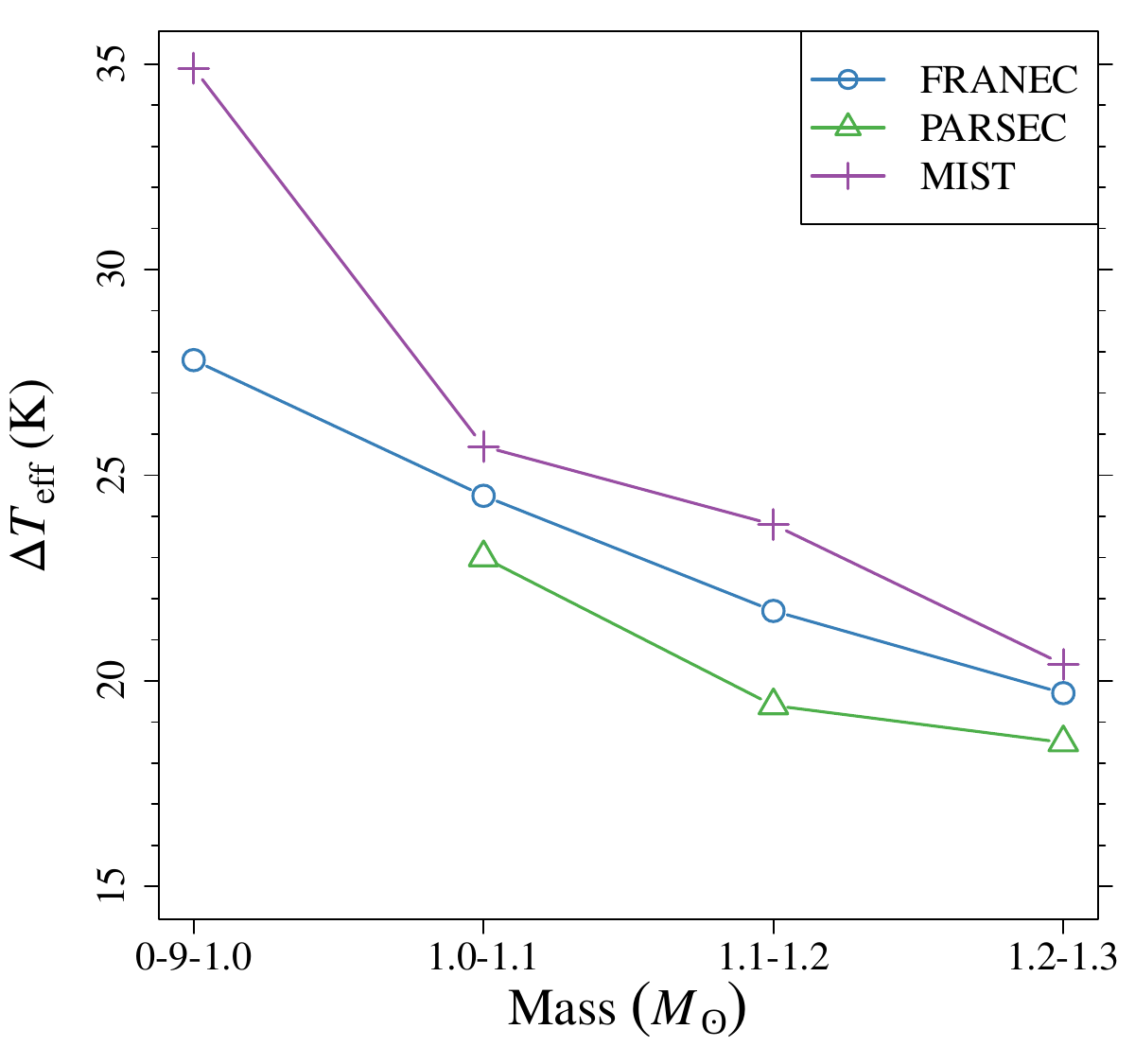}
        \caption{Effective temperature differences between stellar models for a shift  of 0.1 $M_{\sun}$ in the range from 0.9 $M_{\sun}$ to 1.3 $M_{\sun}$. Symbols identify stellar models from different codes. The comparison is performed at [Fe/H] = 0.0, $[\alpha/{\rm Fe}] = 0.0$. }
        \label{fig:dt-cfr-mod}
\end{figure}

The robustness of the  $\Delta T$ dependence on stellar mass
should be further investigated using different stellar evolutionary codes. The  $T_{\rm eff}$ computed by these codes is primarily influenced by the adopted boundary conditions and mixing-length parameter. 
Different choices lead to different offsets as a function of metallicity. This is demonstrated in the analysis presented by \citet{Salaris2018}, who showed the effect of adopting different boundary conditions in the model computations. 
Consequently, effective temperatures from different codes are likely to have distinct zero points and potentially different gradients in the $[\alpha/{\rm Fe}]$--[Fe/H] plane.
However, when all other inputs are held constant, the predicted difference in $T_{\rm eff}$  for a given difference in stellar mass should be highly similar across various codes. This is because this differential effect is not influenced by variations in the adopted $[\alpha/{\rm Fe}]$ and [Fe/H]. 

Figure~\ref{fig:dt-cfr-mod} presents the difference in effective temperature at fixed $\log g=2.75$ for a 0.1 $M_{\sun}$ increase in stellar mass and for baseline masses of 0.9, 1.0, 1.1, and 1.2 $M_{\sun}$. We compare our FRANEC models with MIST models computed using MESA \citep{Choi2016} and PARSEC models \citep{Nguyen2022}\footnote{PARSEC lacks models for  0.9 $M_{\sun}$ in the most recent version of the database.}.
The comparison is made at a common value of $[\alpha/{\rm Fe}]$ = 0.0 and [Fe/H] = 0.0. Unfortunately, PARSEC and MIST databases lack models at different $[\alpha/{\rm Fe}]$, limiting a more comprehensive comparison.

Despite differences in mixing-length parameters, solar reference mixtures, and boundary conditions, the temperature offset and its trend with stellar mass show remarkable similarity across the codes. 
Significant differences arise when comparing these data to those from observations. 
The temperature differences from observed data at $[\alpha/{\rm Fe}]$ = 0.0 and [Fe/H] = 0.0, as resulting from the GAM model, are $6.0 \pm 5.5$ K (0.9-1.0 $M_{\sun}$ shift) and $12.1 \pm 5.2$ K (1.2-1.3 $M_{\sun}$ shift), compared to model predictions of about 30 K and 20 K, respectively. This suggests a significant discrepancy between the trend of APOGEE DR17 $T_{\rm eff}$ with stellar mass and that from models. 
To investigate possible trends induced by the effective temperature calibration in the APOGEE data set, we repeated the analysis using the spectroscopic uncalibrated $T_{\rm eff}$, which revealed an interesting trend. Using spectroscopic data, the difference between 0.9 and 1.0 $M_{\sun}$ (30 K) is in line with the expectation from stellar models. However, significant differences  occur in the following bins, because the difference between uncalibrated effective temperatures decreases to only 7 K in the 1.2 to 1.3 $M_{\sun}$ mass bin.

 \section{Grid-based age estimates}\label{sec:age}

In the presence of the discrepancies between observed and predicted effective temperature, the direct adoption of grid-based estimates of stellar ages is quite problematic.
As discussed, for example, by  \citet{Martig2015} and \citet{Warfield2021}, using the observed $T_{\rm eff}$ to constrain stellar parameters would bias the estimated mass, thus biasing the age. 
Nevertheless, neglecting the $T_{\rm eff}$ constraint is not advisable because, in its absence, grid-based estimates for RGB stars are affected by large uncertainties \citep{Pinsonneault2018}.
A possible remedy \citep{Martig2015, Warfield2021, Warfield2024} is to substitute the effective temperature constraint with that on the stellar mass, as estimated by asteroseismic scaling relations. 
The accuracy of the asteroseismic scaling relations has been extensively investigated and there is evidence of a metallicity, effective temperature, and evolutionary phase-dependent offset in the large frequency separation relation \citep[e.g.][]{White2011, Sharma2016, Stello2022, Li2022}.
Tests carried out on the
scaling relations suggest they are accurate to a level of a few percent 
\citep[e.g.][]{Miglio2016, Rodrigues2017, Brogaard2018, Buldgen2019}.
The approach of fixing the stellar mass to its asteroseismic value is equivalent to adopting the model-predicted effective temperature for the mass given by the scaling relation, $\log g$ being fixed to the observational constraint.
In this section, we adopt this method to estimate the age of the RGB star in our sample.

\begin{table*}[]
\caption{Grid-based estimates of age, mass, and radius for the selected RGB stars.}
\centering
\begin{tabular}{rcccccccccc}
  \hline\hline
 id & apogee & age & age- & age+ & $M$ & $M-$ & $M+$ & $R$ & $R-$ & $R+$ \\ 
    &        &  (Ga) & (Ga) & (Ga) & ($M_{\sun}$) & ($M_{\sun}$) & ($M_{\sun}$) & ($R_{\sun}$) & ($R_{\sun}$) & ($R_{\sun}$) \\
  \hline
1 & 2M13305142-0302596 & 2.28 & 1.12 & 2.05 & 1.48 & 0.26 & 0.33 & 9.99 & 0.62 & 0.73 \\ 
  2 & 2M16593864-2127129 & 12.53 & 4.13 & 1.19 & 0.98 & 0.02 & 0.12 & 4.79 & 0.14 & 0.18 \\ 
  3 & 2M22300191-0652342 & 13.47 & 2.76 & 0.33 & 0.96 & 0.01 & 0.06 & 16.66 & 0.16 & 0.36 \\ 
  \multicolumn{11}{c}{\ldots}\\
  \hline
\end{tabular}
\tablefoot{
The columns contain:
(1) the line number; (2) the APOGEE id; (3-5) the age estimate, with its upper and lower $1 \sigma$ error; (6-8) the mass estimate, with its upper and lower $1 \sigma$ error; (9-11) the radius estimate, with its upper and lower $1 \sigma$ error. Full table is only available in electronic form at the CDS via anonymous ftp to cdsarc.u-strasbg.fr (130.79.128.5) or via http://cdsweb.u-strasbg.fr/cgi-bin/qcat?J/A+A/.}
\end{table*}  

Caution is urged when interpreting age estimates derived from this approach. This method implicitly assumes that unknown factors solely affecting temperature and not evolutionary timescale explain the observed discrepancy in effective temperature. The two primary candidates for such factors are boundary conditions and the mixing-length parameter \citep{Tayar2017, Salaris2018}. Consequently, the reliability of these age estimates depends on the validity of the underlying assumption.

\begin{figure}
        \centering
        \includegraphics[width=8.cm,angle=0]{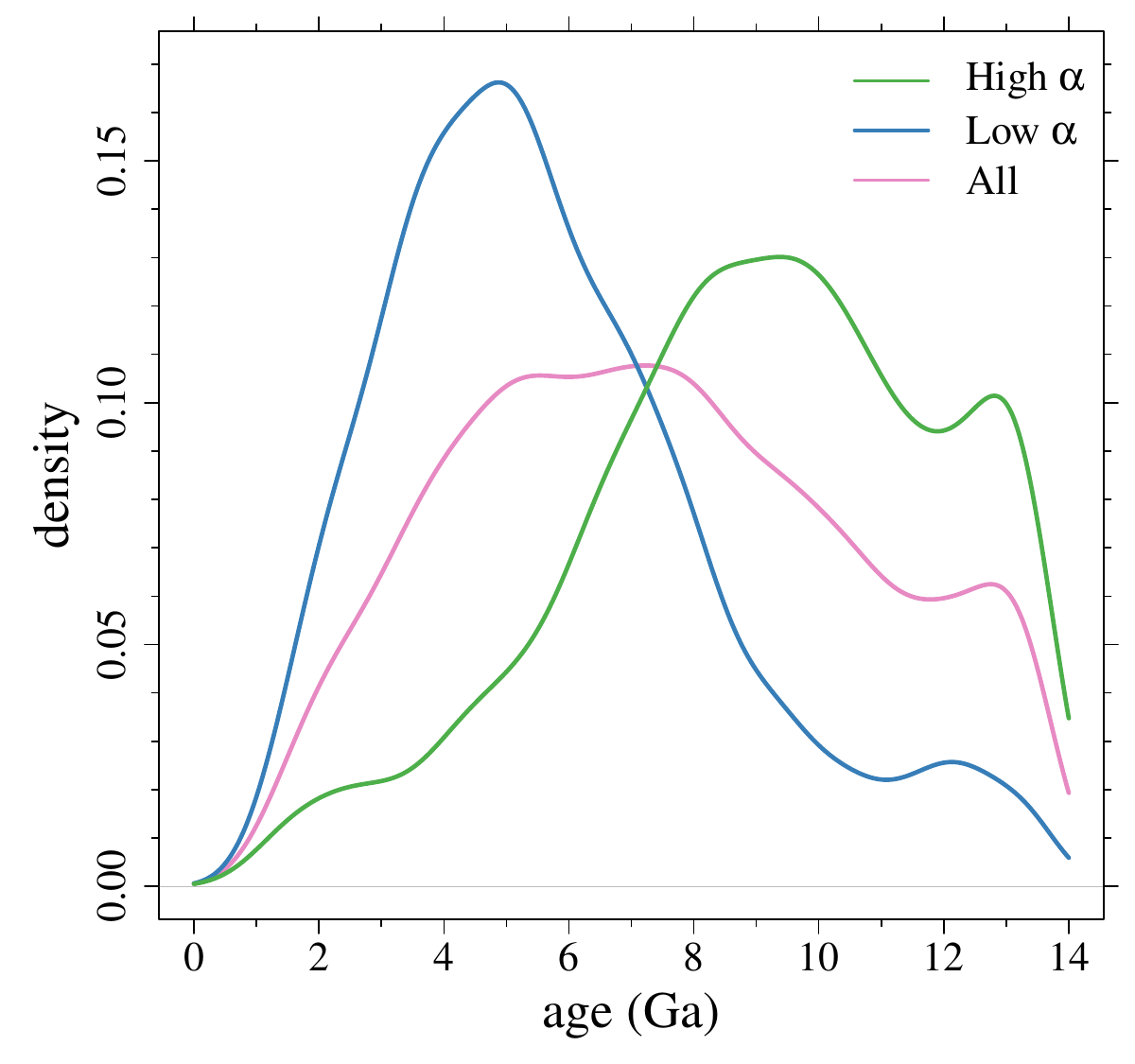}\\
        \includegraphics[width=8.cm,angle=0]{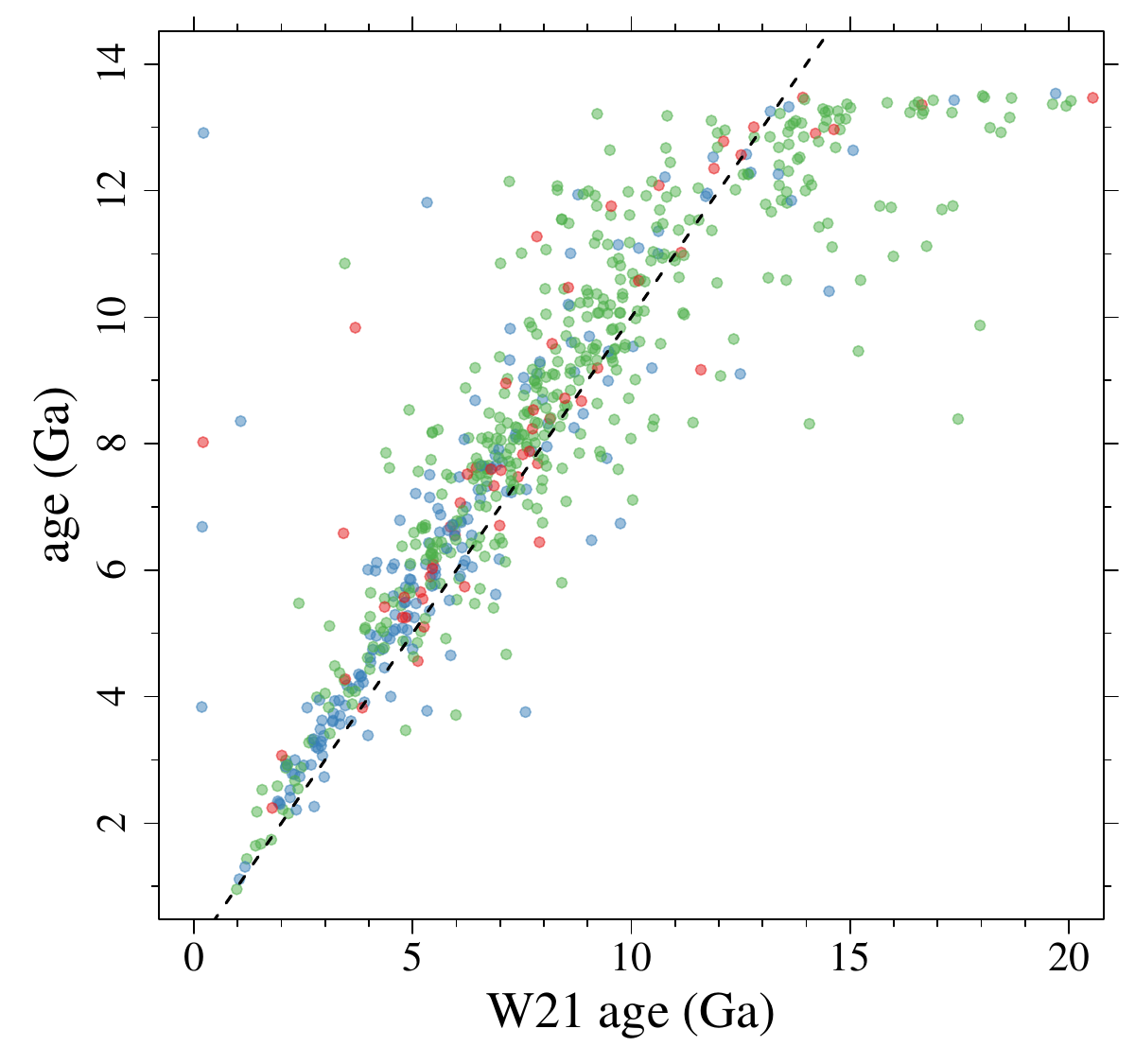}\\
        \includegraphics[width=8.cm,angle=0]{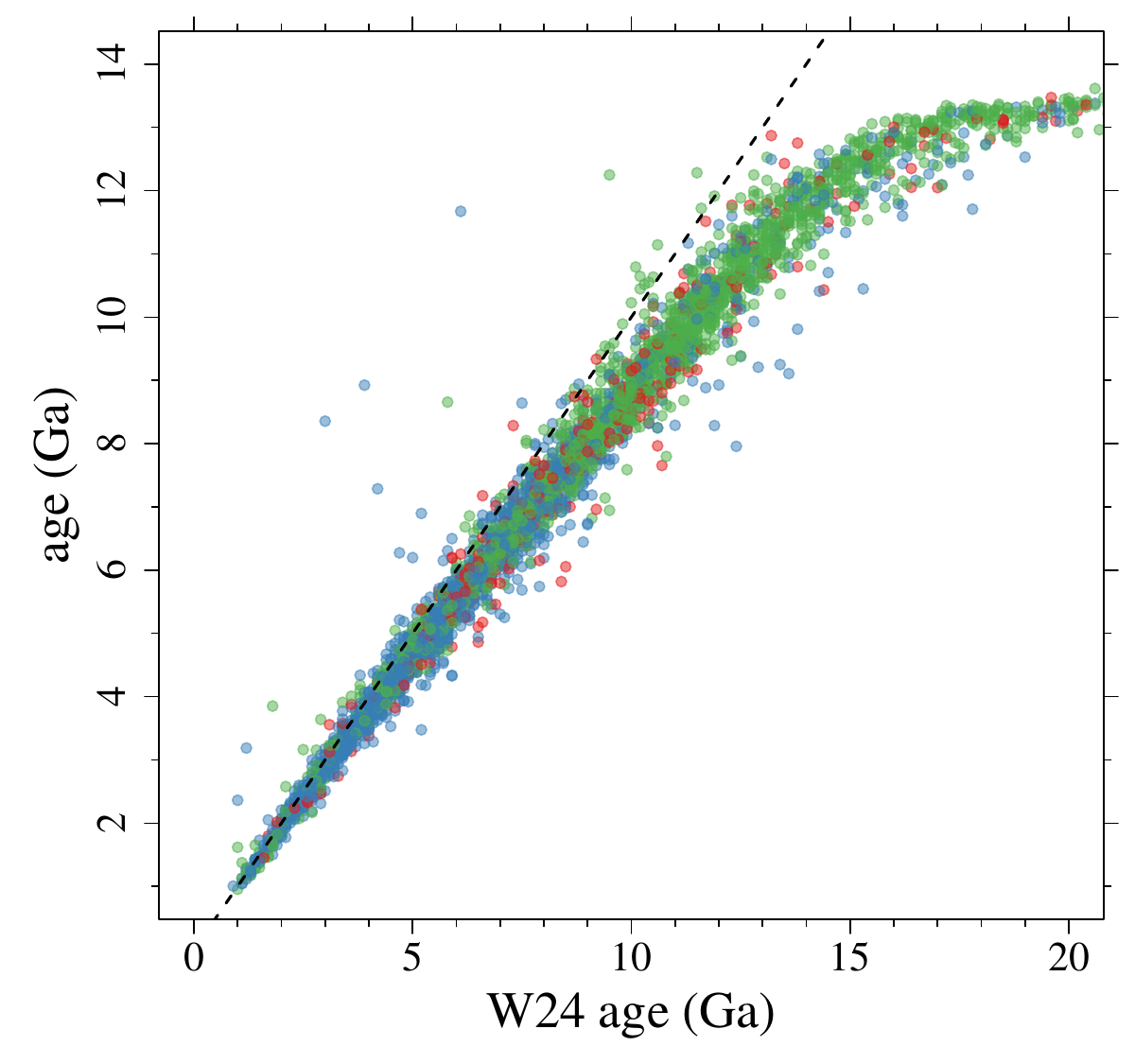}
        \caption{Grid-based estimates of stellar ages. ({\it Top}):  Kernel-density estimators of the age distribution in the full sample (purple line) and in the high- and low-$\alpha$ populations (green and blue lines, respectively).  ({\it Middle}): Comparison of the age estimates from the present paper to those from W21. ({\it Bottom}): Comparison of the age estimates from the present paper to those from W24.}
        \label{fig:age}
\end{figure}

As expected, grid-based estimates of masses and radii are consistent with those from scaling relations. The distribution of relative errors in mass estimates has a median of 0.4\% and  a dispersion characterised by the 16th and 84th percentiles of [-0.7\%, 1.9\%].
The median relative error in radius estimates is 0.4\%  [-0.2\%, 1.8\%]. 
Both distributions show a positive skewness characteristic of RGB grid-based estimates \citep{Gai2011, bulge}.

Figure~\ref{fig:age} presents the age estimates of the stars. Consistent with previous findings, the age distributions between the low- and high-$\alpha$ groups differ significantly (top panel). The low-$\alpha$ group peaks at around 5 Ga, while the high-$\alpha$ population peaks at around 9 Ga. 
The analysis confirms the presence of a significant fraction of young and intermediate-age stars (approximately 6\% younger than 4 Ga and 15\% younger than 6 Ga) in the high-$\alpha$ population, consistent with previous findings \citep[e.g.][]{Chiappini2015, Warfield2021, Zinn2022, Warfield2024}. 
The origin of these stars is still debated. Some authors propose that they are indeed old stars 
that appear young after a merger \citep[e.g.][]{Martig2015,  Jofre2016, Izzard2018, Sun2020, Grisoni2024}, others 
suggest that they are genuinely young and formed in gas relatively unenriched by SNe Ia \citep[e.g.][]{Chiappini2015, Johnson2021} or from bursts of star formation that enhance the
rate of core-collapse supernovae enrichment \citep{Johnson2020}.
Overall, the age distributions in Fig.~\ref{fig:age} closely resemble those found by \citet{Warfield2021} and \citet{Zinn2022}.

The middle and bottom panels of Fig.~\ref{fig:age} show comparisons of our age estimates with those from \citet{Warfield2021} (W21) and \citet{Warfield2024} (W24), respectively. The  comparisons are performed over the samples of common stars, which contain 655 and 4\,246, respectively.  
To enhance readability, error bars are omitted from the figure (the typical $1 \sigma$  error in age is similar for all datasets, being approximately 2 Ga).
The age estimates exhibit good agreement at low age, with only a few stars showing a significantly different prediction. A noticeable disagreement occurs at high ages, possibly due to the difference in the estimation methods. While we impose a cut off of 14 Ga to the models in the grid, no such limit exists in the   W21 and W24 grids. Consequently, in their data set, there are stars with estimated ages of more than 20 Ga. Therefore, this comparison cannot shed light on the importance of the differences in chemical evolution profiles employed in these studies. 
A star-by-star comparison, restricted to stars younger than 14 Ga in both W21 and W24, reveals that our estimates are higher by approximately 0.3 Ga and lower by 0.6 Ga with respect to W21 and W24, respectively. These discrepancies mainly arise for stars older than about 10 Ga, where the different age boundary imposed in the fit becomes relevant.

\begin{figure*}
        \centering
        \includegraphics[width=16.4cm,angle=0]{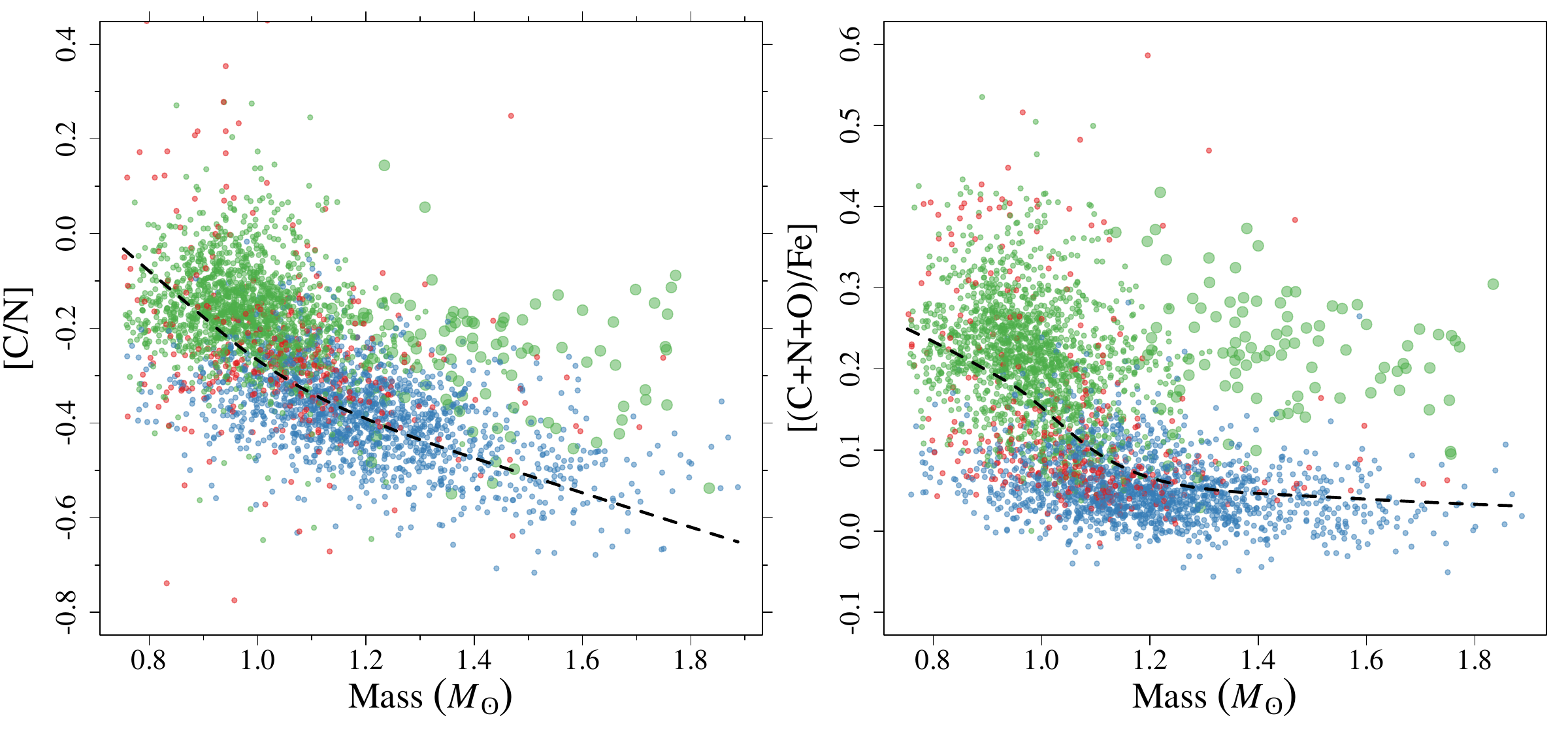}
        \caption{Evolution of surface abundance of different elements with stellar mass. ({\it Left}):  [C/N] ratio evolution. The dashed line shows the predictions from stellar models. The larger symbols identify young $\alpha$-rich stars according to the \citet{Grisoni2024} classification. ({\it Right}): Same as in the left panel but for the [(C+N+O)/Fe] ratio trend. }
        \label{fig:CNO}
\end{figure*}

Despite the global agreement in age estimates, especially for low ages, it is important to note that this is due to a coincidental cancellation effect. 
Important differences exist in the element scaling with $[\alpha/{\rm Fe}]$ between our investigation and those in the literature such as \citet{Martig2015}, \citet{Warfield2021}, \citet{Miglio2021}, and \citet{Warfield2024}. The chemical composition adopted in the present paper for high-$\alpha$ stars becomes progressively O-rich with increasing $[\alpha/{\rm Fe}]$. Simultaneously, the modified composition also impacts the [Fe/H]-to-$Z$ relation. By chance, the two effects have a nearly opposite influence on the stellar evolution timescale, causing age estimates to appear similar even if the underlying assumptions are very different.

Figure~\ref{fig:CNO} shows the  evolution of [C/N] and [(C+N+O)/Fe] ratios with stellar mass. For each star, the surface abundances of C, N, and O  were estimated from a grid-based fit.
The agreement between theoretical models and observations is generally good. The model-predicted [C/N] ratios are systematically lower than the observed values by about 0.04 dex. This difference is less than half of what is expected from variations in computational algorithms used by different stellar evolutionary codes \citep{SilvaAguirre2020}. The discrepancy is even smaller for the predicted and observed [(C+N+O)/Fe] ratios, with a median difference of  0.01 dex.
The figure helps us to investigate on the young population among the high-$\alpha$ stars. These stars were identified as in \citet{Chiappini2015} and \citet{Grisoni2024}, relying on estimated ages and the [Mg/Fe] ratio. Represented by larger symbols in the figure, the high-$\alpha$ young stars clearly deviate from the predicted trends. 
A star-by star comparison reveals that the [C/N] median difference between observations and models is 0.24 dex in the high-$\alpha$ young population, while it is 0.08 dex in the high-$\alpha$ old group. This discrepancy is often attributed 
to mergers or mass-transfer events between two lower-mass giants, which have preserved their original
[C/N] values \citep{Chiappini2015, Jofre2016, Grisoni2024}. Therefore, age estimates for this group are questionable.

\section{Conclusions}\label{sec:conclusions}

Leveraging the recently released APO-K2 catalogue \citep{Stasik2024}, we performed an investigation of the agreement between stellar model predictions and observations for RGB stars. To this purpose, we adopted the element scaling with $[\alpha/{\rm Fe}]$ as resulting from the analysis of \citetalias{Valle2024cno}. The most notable correction is a trend between oxygen and $[\alpha/{\rm Fe}]$ that is steeper than assumed  in a uniform $\alpha$ enhancement scenario.  

The difference in the mixture scaling with $[\alpha/{\rm Fe}]$ impacts stellar models in several ways. First, it alters the evolutionary timescale and the [Fe/H]-to-$Z$ relationship, potentially affecting age estimates from stellar models. Second, it modifies the opacity of stellar matter, consequently influencing the predicted effective temperatures.
These modifications have the potential to significantly change the current paradigm regarding the agreement between stellar models and observations \citep[e.g.][]{Tayar2017, Salaris2018, ML}, and age estimates for high- and low-$\alpha$ stellar populations \citep[e.g.][]{Martig2015, Warfield2021, Miglio2021, Warfield2024}. 

Univariate investigations showed significant differences in the $\Delta T$ offset between data and models with respect to previous investigations by \citet{Tayar2017} and \citet{Salaris2018}. Our analysis does not evidence the $\Delta T$ trend  with [Fe/H] discussed in the literature. Moreover, the $\Delta T$ trend  with stellar mass is reversed with respect to the \citet{Tayar2017} analysis.
A slight difference in $\Delta T$ of 25 K was detected between high- and low-$\alpha$ populations.   

However, these analyses ---which are commonly presented in the literature--- are potentially misleading. The differences in the stellar mass and $\log g$ distributions at different [Fe/H] and $[\alpha/{\rm Fe}]$ values may act as hidden confounding variables. 
To address this issue, we employed a non-linear multivariate statistical model, specifically a generalised additive model  \citep[GAM;][]{Hastie1986, Wood2017} with two smooth terms and a parametric component. This GAM captured the complex relationships between stellar parameters and identified their contributions to $\Delta T$ model-to-data discrepancies.

Our analysis reveals the following key findings.
\begin{itemize}
\item 
The  $\Delta T$ changes with stellar mass, deviating from model predictions. For a mass increase from 0.9 $M_{\sun}$ to 1.2 $M_{\sun}$ (while holding other parameters fixed), the median $\Delta T$ decreases by roughly 55 K. 
This suggests that changes in effective temperature between models at different masses do not correctly reflect observed changes.
\item The $\Delta T$ offset has a strong evolution moving along the $[\alpha/{\rm Fe}]$ axis. At [Fe/H] = $-0.5,$ there is a difference of about 100 K in the effective temperature offset between the extreme $[\alpha/{\rm Fe}]$ zones. 
The $\Delta T$ dependence on  [Fe/H]  is   about one half of that due to $[\alpha/{\rm Fe}]$. 
\item The direction of the steepest $\Delta T$ 
change is almost orthogonal to the line adopted in the separation of high- and low-$\alpha$ populations.
\end{itemize}
Therefore, adopting the scaling of the stellar composition with $[\alpha/{\rm Fe}]$ that mimics the observed trends does not remove the discrepancies between models and predictions; however, it does rule out the possibility that a mismatch in the chemical composition evolution with $[\alpha/{\rm Fe}]$ is the main responsible factor and helps us to assess the contribution of different stellar parameters to this disagreement.  
It is well established that the predicted effective temperatures of RGB stars in stellar models are highly sensitive to the chosen boundary conditions and the mixing-length parameter. These factors have been implicated as potential contributors to the observed discrepancies between theoretical predictions and observed $T_{\rm eff}$ values \citep{Tayar2017, Salaris2018}.
However, a definitive solution to this problem remains elusive. Further theoretical development and alternative approaches might help to reconcile theory and observations  \citep[e.g.][]{Mosumgaard2020, Spada2021, Joyce2023}. 

Due to discrepancies between observed and modelled effective temperatures, using them directly to constrain stellar ages is unreliable. Therefore, we adopted an approach similar to the one by \citet{Martig2015}, \citet{Warfield2021}, and \citet{Warfield2024}. We adopted model-derived effective temperatures for the grid-based age estimation, utilising the stellar mass from the asteroseismic scaling relation to interpolate the model effective temperature.  
The resulting age estimates are consistent with those reported in the literature \citep[e.g.][]{Martig2015, Warfield2021, Miglio2021, Warfield2024}, confirming the presence of a 6\% fraction of stars younger than 4 Ga in the high-$\alpha$ population. A star-by-star comparison with W21 reveals a median difference of 0.4 Ga, with our estimates generally higher. The same comparison with W24 shows a median difference of -0.6 Ga for stars younger than 14 Ga in both data sets. Analyses of the [C/N] ratio support a possible origin of the young high-$\alpha$ population in mergers or mass-transfer events \citep{Chiappini2015, Grisoni2024}.

While our estimated ages globally agree with previous results, this agreement comes from a serendipitous cancellation effect. Our investigations rely on  stellar models computed under different chemical abundances and input physics. Notably, the chemical composition adopted in our analysis  becomes progressively O-rich increasing $[\alpha/{\rm Fe}]$ with respect to a standard $\alpha$ enhancement scenario. This also 
modifies the [Fe/H]-to-$Z$ relationship. 
By chance, the two effects have  a nearly opposite impact on the stellar evolution timescale, resulting in very similar age estimates even if the underlying assumptions are very different. 
The  resilience of age estimates to changes in the scaling between chemical abundance and $[\alpha/{\rm Fe}]$ bolsters our confidence in the  findings of the existing literature.

\begin{acknowledgements}
G.V., P.G.P.M. and S.D. acknowledge INFN (Iniziativa specifica TAsP) and support from PRIN MIUR2022 Progetto "CHRONOS" (PI: S. Cassisi) finanziato dall'Unione Europea - Next Generation EU.
\end{acknowledgements}

\bibliographystyle{aa}
\bibliography{allbib}

\appendix

\section{Analysis of GAM residual}\label{app:gam}

\begin{figure}[ht!]
        \centering
        \includegraphics[width=8.2cm,angle=0]{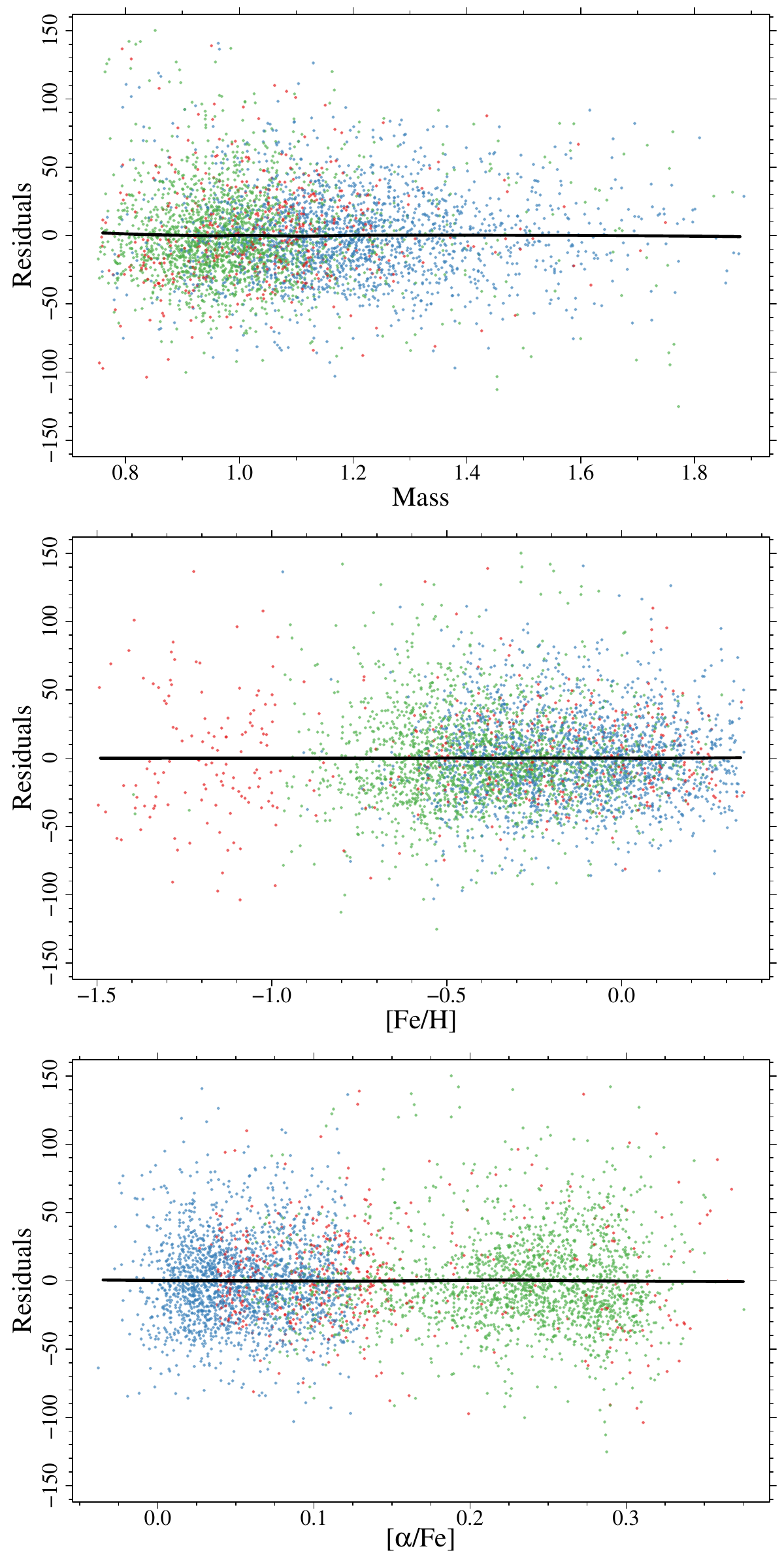}
        \caption{GAM residuals as a function of the stellar mass (in solar units), the metallicity [Fe/H], and $[\alpha/{\rm Fe}]$. The black lines are the smoother of the residuals. }
        \label{fig:gam-resid}
\end{figure}

This section report the analysis of the adequacy of the GAM models in Eq.~(\ref{eq:gam}), adopted in the paper.
The first check which should be performed for a GAM model is the adequacy of the basis dimension for the smooth terms.
This test is based on computing an estimate of the residual variance based on differencing residuals that are near neighbours according to the covariates of the smooth \citep{Wood2017}. This estimate, divided by the residual variance, is the $k$-index. The $p$-value is computed by simulation \citep{Wood2017}. The results of the test are collected in Tab.~\ref{tab:k'}. The hypothesis of the adequacy of the base dimension cannot be rejected.

\begin{table}[h]
    \centering
    \caption{Adequacy test for the basis dimension for the two  smooth terms.}
    \label{tab:k'}
    \begin{tabular}{lcccc}
    \hline\hline
     term & $k'$ &   edf & $k$-index & $p$-value \\
    \hline                       
$f_1({\rm[Fe/H]} , [\alpha/{\rm Fe}] , \log g)$   &189  &85.14    &1.01    &0.78\\
 $f_2(M)$                            &14   &5.79    &0.99    &0.37\\
\hline
    \end{tabular}
    \tablefoot{$k'$ is the dimension of the basis minus 1; edf are the effective degree of freedom of the smooth; $k$-index is the test statistic; $p$-value provides the statistical significance of the test. }
\end{table}

The second test is the analysis of deviance residuals, to verify that the model correctly captured all the trends present in the data.  Figure~\ref{fig:gam-resid} shows the model residuals as a function of the stellar mass, the metallicity [Fe/H], and $[\alpha/{\rm Fe}]$, the smoother of the residuals, shown in the panels, demonstrate that no unaccounted trends are present in the model residuals.

The analysis of the model residuals shows the presence of 38 stars with residuals above 100 K that is, about $3 \sigma$ above the mean. This high prevalence of outliers at high effective temperature suggests that these stars may possibly be RC stars, misclassified in the RGB group. In fact, the classification in the APO-K2 catalogue does not rely on robust asteroseismic detection of the period spacing but only on discriminating functions based on [C/N], $\log g$, and $T_{\rm eff}$. 
However, the removal of these stars from the sample adopted for the GAM fit only marginally improve its performance and predictive capabilities. Therefore in the analysis we adopt the models computed without rejecting these potential outliers.

\end{document}